\documentclass[hyper,11pt,paper]{article}
\usepackage{jheppub}

\usepackage[usenames,dvipsnames,table]{xcolor}
\usepackage{graphicx,amsmath,amssymb,amsthm,multirow,array,bm,bbm,esint}
\usepackage[mathscr]{eucal}
\usepackage[bbgreekl]{mathbbol}
\usepackage{epsf,amsfonts}
\usepackage{hyperref}
\usepackage{slashed}
\usepackage[normalem]{ulem}

\newcommand{\beq}{\begin{equation}}
\newcommand{\eeq}{\end{equation}}
\newcommand{\M}{\mathcal{M}}

\title{'t Hooft anomalies and boundaries}

\author[a]{Kristan Jensen,}
\author[b]{Evgeny Shaverin,}
\author[c]{and Amos Yarom}

\affiliation[a]{Department of Physics and Astronomy, San Francisco State University, San Francisco, CA 94132}
\affiliation[b]{Department of Physics, Technion, Haifa 32000, Israel}

\emailAdd{kristanj@sfsu.edu}
\emailAdd{evgeny@tx.technion.ac.il}
\emailAdd{ayarom@physics.technion.ac.il}

\abstract{
We argue that there is an obstruction to placing theories with 't Hooft anomalies on manifolds with a boundary, unless the symmetry associated with the anomaly can be represented as a non-invariance under an Abelian transformation. For a two dimensional conformal field theory we further demonstrate that all anomalies except the usual trace anomaly are incompatible on a manifold with a boundary. Our findings extend a known result whereby, under mild assumptions, Lagrangian theories with chiral matter cannot be canonically quantized.  
}

\preprint{\today}

\begin{document}

\maketitle

\section{Introduction}

Symmetries play a fundamental role in characterizing quantum field theories. When a Lagrangian description is available, continuous symmetries are often in one-to-one correspondence with Noether currents which are conserved inside correlation functions. If a continuous symmetry of the Lagrangian does not correspond to a conserved current we say that the symmetry is anomalous. A prime example of an anomalous symmetry is the $U(1)_A$ symmetry of the Standard model, whereby the axial current is not conserved. In this work we will consider 't Hooft anomalies in which the non-conservation law is characterized by a $c$-function.

Even when a Lagrangian description is not available one can characterize symmetries by an invariance of the generating function of connected correlators, $W$, under a transformation, $\delta_v$  of the external fields. If $\delta_v$ is a symmetry then one may associate an algebra with it. A set of transformations $\delta_v$ which satisfy an appropriate algebra but under which $\delta_v W$ is nonzero are said to generate an anomalous symmetry. If $\delta_v W$ is a nonzero, local, functional of the external fields, then the theory is said to possess an 't Hooft anomaly.\footnote{There may also be 't Hooft anomalies for discrete symmetries, like the time-reversal (or parity) anomaly in three dimensions. We do not consider anomalies for discrete symmetries in this work.}

Anomalies serve as one of the few probes of non-perturbative quantum field theories. Anomaly matching between fixed points of RG flow \cite{1980} allows one to posit the existence of various dualities \cite{Seiberg:1994bz}, it allows for the computation of anomalous couplings of D-branes \cite{Green:1996dd} and for understanding black hole entropy \cite{Kraus:2005vz}. From a theoretical standpoint the classification of anomalies leads to a rich cohomological structure \cite{Wess:1971yu}, and to index theorems \cite{Atiyah1963}. Novel studies have also tied anomalies to transport \cite{Son:2009tf,Golkar:2012kb,Jensen:2012kj,Jensen:2013kka,Jensen:2013rga,Golkar:2015oxw,Chowdhury:2016cmh} leading to new experimental signatures of anomalies in condensed matter systems \cite{Li:2014bha,Gooth:2017mbd} and possibly astrophysical settings \cite{Kaminski:2014jda,Shaverin:2014xya}.

While much is known of anomalies on manifolds without boundaries, less is known regarding anomalies on manifolds with boundaries. There is some evidence that chiral fermions can not be canonically quantized on a manifold with boundary \cite{Luckock:1990xr}. For free fermions subject to boundary conditions which preserve Lorentz invariance along the boundary and can be implemented mode by mode, the boundary conditions are incompatible with the chirality projection operator. Thus there are no linear, local, Lorentz-invariant boundary conditions for chiral fermions. We reproduce the relevant results of  \cite{Luckock:1990xr} in appendix \ref{A:canonical}.

Alternate versions of the arguments in \cite{Luckock:1990xr} involve placing constraints on the value of the stress tensor near the boundary \cite{McAvity:1993ue}. These allow extensions of the results of \cite{Luckock:1990xr} to theories of self-dual $p$-forms in $d=2p+2$ dimensions and to include bulk interactions \cite{Herzog:2017xha}. However, such arguments still rely on a Lagrangian description, linear and local boundary conditions, and no couplings between bulk and boundary degrees of freedom. One can ask if, upon relaxing these assumptions, there remains an obstruction to quantization. In $d\geq 4$, it is easy to devise local but nonlinear boundary conditions for chiral fermions which preserve all of the symmetries of the classical theory. For instance, a candidate boundary condition for a $d=4$ Weyl fermion $\psi$ is that $\bar{\psi} n^{\mu}\gamma_{\mu}\psi$ vanishes on the boundary, with $n^{\mu}$ the normal vector to the boundary. We are unaware of an argument that precludes quantization of chiral matter subject to such boundary conditions.

In this note we endeavor to obtain fully non-perturbative constraints on theories with 't Hooft anomalies, such as theories of chiral matter. These constraints do not rely on linear boundary conditions, nor on a Lagrangian description. Our chief assumption is locality, even at the boundary. Our main result is a proof that there is an obstruction to placing theories with anomalous symmetries on a manifold with a boundary, unless the anomaly can be represented as the non-invariance under an Abelian transformation (anomalies of this type are sometimes called Abelian). Our proof relies on the Wess-Zumino consistency condition \cite{Wess:1971yu}, which is known to provide rather weak constraints on Abelian anomalies. Indeed, using Ward identities and conformal invariance, one can demonstrate that a two-dimensional conformal field theory (CFT) on the upper half plane can not possess an Abelian anomaly. Put differently, two dimensional CFT's must have equal central charges and equal Kac-Moody levels, viz., $c_L=c_R$ and $k_L=k_R$. That on the half plane $c_L=c_R$ was demonstrated in \cite{Billo:2016cpy}. In our work we rederive and strengthen this result and extend it to more general anomalies.

Our main proof that the Wess-Zumino consistency condition allows only Abelian anomalies on a manifold with a boundary can be found in Section \ref{S:anomalies}. Since the brunt of the proof is somewhat formal we provide several explicit examples in a more hands-on manner: in Subsection \ref{SS:2d} we discuss in detail all possible 't Hooft anomalies in two dimensions. We show explicitly that two dimensional flavor anomalies are not compatible on a manifold with a boundary but gravitational anomalies are. We also discuss two dimensional Weyl anomalies and Lorentz-Weyl anomalies. The latter suffers from an obstruction similar to that of non-Abelian anomalies. Four and higher dimensions are analyzed in Subsections \ref{SS:4d} and \ref{SS:higherd}. Our argument that two-dimensional CFT's with a boundary do not support even Abelian anomalies is given in Section~\ref{S:2dCFT}. We end with a brief discussion of our results in Section \ref{S:discussion}.

\emph{Note}: Some time ago, we learned of related work by S.~Hellerman, D.~Orlando, and M.~Watanabe~\cite{watanabe,simeon}. They also find obstructions to placing theories with anomalies on spaces with boundary. Their work agrees with ours where they overlap.

\section{Anomalies}
\label{S:anomalies}

A classical symmetry is a transformation of the dynamical fields under which the action remains invariant. In quantum field theory a symmetry manifests itself as an invariance of the generating function $W[A]$ under a transformation the external sources, for instance
\begin{equation}
\label{E:Symmetry}
	\delta_{v} W[A] = W[A+\delta_{v} A] - W[A] = 0\,,
\end{equation}
where $\delta_{v }A$ is an infinitesimal transformation parameterized by a set of functions $v$. Let us consider a particular class of transformations, 
\begin{equation}
\label{E:smalldelta}
	\delta_{v}A = d v + [A,v] \equiv D v \,,
\end{equation}
After integration by parts, and in the absence of a boundary, we have
\begin{equation}
\label{E:deltav}
	\delta_{v} W  =  - \int_{\mathcal{M}} d^dx \,v \,D_{\mu}\left(  \frac{\delta  W}{\delta A_{\mu}}\right)\,.
\end{equation}
In \eqref{E:deltav} we have removed all but a spacetime index for brevity.
Thus,
\begin{equation}
\label{E:WZconsistency}
	[\delta_{v_1},\,\delta_{v_2}] W = \delta_{[v_1,\,v_2]} W
\end{equation}
which is trivially satisfied when $A \to A + \delta_{v}A$ is a symmetry \eqref{E:Symmetry}. If there exists a non-trivial solution to \eqref{E:WZconsistency} for which $\delta_{v} W$ is a local function of the external fields then we say that the symmetry \eqref{E:Symmetry} is anomalous. By a non-trivial solution we mean a solution which does not reduce to $G=0$ upon adding to the quantum action terms which are local in $A$ or its derivatives.

Equation \eqref{E:WZconsistency} is often written in a slightly different form. Denoting the anomalous variation of $W$ as
\begin{equation}
\label{E:deltav2}
	\delta_{v} W[A] =  - \int_{\mathcal{M}} v G\,,
\end{equation}
with $G$ a volume form and assigning $\delta_{v_2}v_1=0$, \eqref{E:WZconsistency} amounts to
\begin{equation}
\label{E:WZwithG}
	0=
	-\int_{\mathcal{M}} \left( v_2 \delta_{v_1} - v_1 \delta_{v_2} - [v_1,\,v_2] \right) G\,.
\end{equation}

Non trivial solutions to \eqref{E:WZwithG} exist in even dimensional spacetimes. They can be obtained by converting \eqref{E:WZwithG} to a cohomological problem. While this is textbook material \cite{Zumino:1983ew,AlvarezGaume:1984dr,bertlmann} it is instructive to sketch the conversion procedure in some detail. The integrated version of the transformation specified by \eqref{E:smalldelta} amounts to
\begin{equation}
\label{E:largedelta}
	A \to \bar{A} = g^{-1}(A+d)g\,.
\end{equation}
Indeed, the transformation \eqref{E:largedelta} reduces to \eqref{E:smalldelta} for $g=1+v+\mathcal{O}(v^2)$, expanded to linear order in $v$. Let us consider transformations $g$ which depend on parameters $\theta^{\alpha}$, which take values on the two-sphere $\mathbb{S}^2$, such that $g(x,\,\theta)\big|_{\theta=0}=1$. 
Note that $\bar{A}(x,\theta) \big|_{\theta=0} = A(x)$. 

One reason for considering a two dimensional parameter space is that we will be associating the two transformations, $v_1$ and $v_2$, appearing in \eqref{E:WZwithG} with each of the two angles. (The authors of \cite{Zumino:1983ew,AlvarezGaume:1984dr,bertlmann} consider a parameter space which is a $p$-sphere to obtain a more general result than needed in this summary.)  To wit, consider
\begin{equation}
	v_{\alpha}(x,\theta) = g^{-1}(x,\,\theta) \frac{\partial}{\partial \theta^{\alpha} }g(x,\theta)
\end{equation}
such that $v_1(x)$ and $v_2(x)$ from \eqref{E:WZconsistency} are the $\theta \to 0$ limit of $v_1(x,\,\theta)$ and $v_2(x,\,\theta)$. The strategy for solving \eqref{E:WZconsistency} is to solve it for arbitrary $\theta$ and then restrict the solution to $\theta=0$.

To proceed, it is convenient to consider the gauge parameters $v_{\alpha}$ as a one-form in parameter space, 
\begin{equation}
	\hat{v} = v_{\alpha} d\theta^{\alpha} = g^{-1} \hat{d} g
\end{equation}
where $d\theta^{\alpha}$ is a basis of one-forms on the $\mathbb{S}^2$ and $\hat{d}$ is the exterior derivative on $\mathbb{S}^2$. Thus, we have, for instance,
\begin{equation}
\label{E:identity1}
	\hat{d} \hat{v} = -\hat{v} \wedge \hat{v} \,.
\end{equation}
If we define $\bar{F} = d\bar{A} + \bar{A} \wedge \bar{A}$ then
\begin{equation}
\label{E:hatdisdelta}
	\hat{d} \bar{A} = -d \hat{v} - \left( \bar{A} \wedge \hat{v} + \hat{v} \wedge \bar{A} \right) \qquad \hat{d} \bar{F} = [\bar{F},\,\hat{v}] \,.
\end{equation}  
Equation \eqref{E:hatdisdelta}  implies that we can make the replacement
\begin{equation}
\label{E:hatisdelta}
	\hat{d} = (d\theta^{\alpha}) \delta_{v_{\alpha}}\,,
\end{equation}
whenever $\hat{d}$ acts on $\bar{A}$ or $\bar{F}$ (but not on $\hat{v}$). 

Extending \eqref{E:WZwithG} to $\mathcal{M} \times \mathbb{S}^2$ and contracting with $d\theta^1 \wedge d\theta^2$ we find, using \eqref{E:identity1}, that \eqref{E:WZwithG} amounts to
\begin{equation}
\label{E:WZconsistencyB}
	0= -\int_{\mathcal{M}\times \mathbb{S}^2} \hat{d} \left(\hat{v} G \right)\,.
\end{equation}
Equation \eqref{E:WZconsistencyB} may be solved if there exists a $Q$ such that
\begin{equation}
\label{E:descent}
	\hat{d} \left(\hat{v} G\right) = d Q\,.
\end{equation}
In fact, there is an entire equivalence class of solutions to \eqref{E:WZconsistencyB}. If $\hat{v}G$ solves \eqref{E:descent} then so does $\hat{v}G+\hat{d}\hat{G} + dG_b$. We write $\hat{v}G \sim \hat{v}G+\hat{d}\hat{G} + dG_b$. From a physical perspective, the $\hat{d}\hat{G}$ term amounts to adding a trivial solution to \eqref{E:WZconsistency}. Given that $d\hat{d} + \hat{d}d=0$, the $d G_b$ corresponds to adding a boundary term to $W$ which is local in $A$ or its derivatives. There is a similar equivalence class for $Q$. 

Equation \eqref{E:descent} can be solved using the celebrated descent relations, (see, for instance, \cite{bertlmann}). A particularly useful class of solutions is given by \cite{Zumino:1984ws} 
\begin{equation}
\label{E:vGsolution}
	\hat{v} G \sim (m+1)m \int dt (1-t) P(d\hat{v},\,\bar{A},\,\bar{F}_t^{m-1} )\,, \quad m \geq 1 
\end{equation}
and 
\begin{equation}
\label{E:Qsolution}
	Q  \sim \begin{cases}
	\frac{1}{2} (m+1)m(m-1) \int dt (1-t)^2 P(d\hat{v}^2,\,\bar{A},\,\bar{F}_t^{m-2}) \,, 
	& m \geq 2 \\
	\hbox{Tr} \left(\hat{v} d\hat{v} \right)  & m=1
	\end{cases}
\end{equation}
in $d=2m$ spacetime dimensions. Here we have defined $\bar{F}_t = t d \bar{A} + t^2 \bar{A} \wedge \bar{A}$ and $P$ is the $d+2$ dimensional anomaly polynomial.  Given that $P = \hbox{Tr} \left( F_1 \wedge F_2 \wedge G \right)$ with $F_1$ and $F_2$ two form field strengths and $G$ a $d-2$ form, then $P(X,Y,Z) = \hbox{Tr} \left(X \wedge Y \wedge Z\right)$. Equation \eqref{E:descent} and its solution \eqref{E:vGsolution} and \eqref{E:Qsolution} are the  results we need from \cite{Zumino:1983ew,AlvarezGaume:1984dr,bertlmann}.

In the presence of a boundary the functional form of $\delta_v$ \eqref{E:deltav} will receive boundary contributions, so that
\begin{equation}
	\delta_v W[A] = \int_{\partial \mathcal{M}\times \mathbb{S}^2} v G_b - \int_{\mathcal{M}\times \mathbb{S}^2} v G\,.
\end{equation}
The previous analysis will go through as before such that  \eqref{E:WZconsistencyB} takes the form
\begin{equation}
	0 = \int_{\partial \mathcal{M}\times \mathbb{S}^2} \hat{d} \left( \hat{v} G_b \right) - \int_{\mathcal{M}\times \mathbb{S}^2} \hat{d} \left( \hat{v} G \right)\,.
\end{equation}
In order to avoid setting $\hat{v} G = d \left(\hat{v} G_b\right) + \hat{d} G'$ (which implies a trivial solution to the Wess Zumino consistency conditions, $\delta_v W = \int \hat{d} G'$) we must first look for a solution of the form \eqref{E:descent}, and then set
\begin{equation}
\label{E:boundaryremainder}
	\hat{d}  \left( \hat{v} G_b \right) = Q\,.
\end{equation}
From a physical standpoint \eqref{E:boundaryremainder} implies that the boundary terms generated by the standard bulk anomaly have to be compensated for by additional boundary terms in order that the Wess-Zumino conistency condition is satisfied in the bulk and the boundary. This is a standard technique which has been used to classify Weyl anomalies on manifolds with boundaries in, e.g., \cite{polchinski1998string,Herzog:2015ioa}.

By extending the parameter space for the transformations $g(x,\theta)$ from $\mathbb{S}^2$ to $\mathbb{S}^3$, a necessary condition for \eqref{E:boundaryremainder} to be satisfied is that
\begin{equation}
\label{E:hatdQ0}
	\hat{d}Q=0\,. 
\end{equation}
Acting with $\hat{d}$ on the solution given in \eqref{E:Qsolution} we find that in four and higher spacetime dimensions $\hat{d}Q\neq 0$ unless $Q=0$. The condition $Q=0$ is satisfied if $d\hat{v}^2 = d\theta^1 d\theta^2 [dv_1,\,dv_2]=0$ which implies that the symmetry is Abelian. In two dimensions $\hat{d}Q=0$ only for an Abelian symmetry. As we will show shortly in that case \eqref{E:boundaryremainder} can, indeed, be satisfied. Thus, we conclude that 't Hooft anomalies on manifolds with boundaries are consistent only as long as the anomaly polynomial takes the form 
\beq
P = F\wedge \hdots\,,
\eeq
with $F$ the curvature of an Abelian connection. 

Anomalies of this type may be represented as a non-invariance under an Abelian transformation. These include pure Abelian anomalies, like a $U(1)^3$ anomaly in four dimensions, or anomalies which are mixed between an Abelian symmetry and a non-Abelian one. This result also applies to $SO(d)$ Lorentz anomalies: in two dimensions, the $SO(2)$ Lorentz anomaly is consistent on a space with boundary, but there is an obstruction for pure Lorentz anomalies in more than two dimensions.

Thus, we conclude that the Wess-Zumino consistency condition cannot be satisfied for theories with non-Abelian 't Hooft anomalies on manifolds with boundaries. In the remainder of this section we will study the ramifications of this result in two, four, and higher dimensions.

\subsection{Two dimensions}
\label{SS:2d}

In order to illustrate our main result, consider flavor anomalies in two-dimensional field theories. Let $A$ be a connection associated with a non-Abelian transformation, such that
\begin{align}
\begin{split}
\label{E:rules1}
	\delta_{\Lambda}A = d \Lambda + [A,\Lambda]\,,
	\qquad
	\delta_{\Lambda_1} \Lambda_2 = 0
	\,.
\end{split}
\end{align}
Equation \eqref{E:vGsolution} reads
\begin{align}
\begin{split}
\label{E:WZsolNA}
	\delta_{\Lambda} W &= -k \int_{\mathcal{M}} \hbox{Tr} \left( d\Lambda A \right)  \\
		&=-k\int_{\partial \mathcal{M}} \hbox{Tr} \left(\Lambda A\right)  + k \int_{\mathcal{M}} \hbox{Tr} \left( \Lambda dA \right) 
\end{split}
\end{align}
where we have integrated by parts in the second line.
A direct computation gives
\begin{equation}
\label{E:nonWZNA}
	[\delta_1,\,\delta_2]W - \delta_{[1,2]}W = 	k  \int_{\partial \mathcal{M}} \hbox{Tr} \left(\Lambda_1 d\Lambda_2 - \Lambda_2d\Lambda_1\right) \,.
\end{equation}
which matches \eqref{E:Qsolution}. Thus, as expected, in the absence of a boundary, \eqref{E:WZsolNA} solves the Wess-Zumino consistency condition. 

In the presence of a boundary, we may add to \eqref{E:WZsolNA} a boundary variation $\delta_\Lambda W = k\int_{\partial \mathcal{M}} \hbox{Tr}(\Lambda A)$ such that \eqref{E:nonWZNA} takes the form
\begin{equation}
\label{E:2dfinal}
	[\delta_1,\,\delta_2]W - \delta_{[1,2]}W = 	k  \int_{\partial \mathcal{M}} \hbox{Tr} \left( [\Lambda_1,\,\Lambda_2]A \right)
\end{equation}
which vanishes only when the symmetry is Abelian. 

As stated in our general discussion, the boundary terms \eqref{E:2dfinal} can not be removed by further adding local boundary contributions to \eqref{E:WZsolNA}. Indeed, in the presence of a boundary one may attempt to modify \eqref{E:WZsolNA} to,
\begin{equation}
\label{E:plusb}
	\delta_{\Lambda}W = k \int_{\mathcal{M}} \hbox{Tr} \left(\Lambda d A \right) 
	+ b \int_{\partial \mathcal{M}} \hbox{Tr} \left( \Lambda A \right)\,. 
\end{equation}
One may check that $[\delta_1,\delta_2]W-\delta_{[1,2]}W$ does not vanish for any non zero $b$. It may be the case that non-local expressions may be added to the right hand side of \eqref{E:plusb} so that the right hand side of \eqref{E:nonWZNA} vanishes. While possible it would mean that the non-conservation law for the anomalous current would involve a non-local expression.

Let us move onward to two-dimensional theories with a gravitational anomaly. Recall that gravitational anomalies may manifest themselves as Lorentz anomalies or Einstein anomalies. For general dimensions, a Lorentz transformation $\delta_\theta$  of the vielbein $e^a_{\mu}$ and spin connection $\omega_{\mu}{}^{a}{}_{b}$ are given by 
\begin{equation}
\label{E:Lorentztransformationini}
	\delta_{\theta} e^a_{\mu} = -\theta^a{}_b e^b{}_{\mu}\,,
	\quad
	\delta_{\theta} \omega_{\mu}{}^a{}_{b} = \omega_{\mu}{}^a{}_c \theta^c{}_b - \theta^a{}_c \omega_{\mu}{}^c{}_b + \partial_{\mu} \theta^{a}{}_b\,.
\end{equation}
In two dimensions the Lorentz group is Abelian and we may define $\theta^{a}{}_{b} = \theta \epsilon^{a}{}_{b}$ and $\omega_{\mu}{}^{ab} = \omega_{\mu} \epsilon^{ab}$ with $\epsilon^{ab}$ the Levi-Civita tensor. It is also common to define the spin connection as a one form, $\omega = \omega_{\mu}dx^{\mu}$. In this language \eqref{E:Lorentztransformationini} become
\begin{equation}
\label{E:Lorentztransformation}
	\delta_{\theta} e^a{}_{\mu} = -\theta \epsilon^a{}_b e^b{}_{\mu}
	\qquad
	\delta_{\theta} \omega = d \theta\,.
\end{equation}
Likewise, Einstein (diffeomorphism) transformations of the vielbein $e^a{}_{\mu}$, metric $g_{\mu\nu}$ and Christoffel connection one-form $\Gamma^\mu{}_\nu = \Gamma^{\mu}{}_{\nu\alpha}dx^{\alpha}$ are given by
\begin{align}
\begin{split}
\label{E:geometry1}
	\delta_{\xi} g_{\mu\nu} &= \partial_{\mu}\xi^{\rho} g_{\rho\nu} + \partial_{\nu} \xi^{\rho} g_{\mu\rho} \\
	\delta_{\xi} \Gamma^{\mu}{}_{\nu} &= \partial_{\nu} \xi^{\rho} \Gamma^{\mu}{}_{\rho} - \partial_{\rho} \xi^{\mu} \Gamma^{\rho}{}_{\nu} + d \partial_{\nu} \xi^{\rho} \\
	\delta_{\xi} e_{\mu}^a &= \partial_{\mu}\xi^{\nu}e_{\nu}^a 
	\\
	\delta_{\xi} \omega &= 0 \,.
\end{split}
\end{align}
In addition $\xi$ transforms as a tangent vector, and $\theta$ as a scalar, viz.
\begin{align}
\begin{split}
\label{E:geometry2}
	\delta_{\xi_1} \xi_2^{\mu} & = -(\partial_{\rho} \xi_1^{\mu})\xi_2^{\rho}\,,
	\qquad
	\delta_{\xi} \theta = 0\,, \\
	\delta_\theta \theta_1 &= 0 \,,
	\hspace{0.83in}
	\delta_\theta \xi^{\mu} = 0\,.
\end{split}
\end{align}
We are using the so-called ``passive'' representation of diffeomorphisms in~\eqref{E:geometry1} and~\eqref{E:geometry2}. Their active counterparts will not be covered in this work.

The solution to the Wess-Zumino consistency condition on a manifold without a boundary for two-dimensional Lorentz anomalies can be read off of \eqref{E:vGsolution}
\begin{equation}
\label{E:LorentzAnomaly}
	\delta_v W = -\tilde{c} \int_{\mathcal{M}} d\theta\, \omega \,,
	\qquad
	\delta_\xi W = 0\,.
\end{equation}
The transformation properties of $\omega$ \eqref{E:Lorentztransformation} and the Lorentz variation of $W$ \eqref{E:LorentzAnomaly} are identical to those of the flavor anomaly \eqref{E:rules1} and \eqref{E:WZsolNA} upon identifying the spin connection with an Abelian gauge connection. Thus, from \eqref{E:nonWZNA} and the Abelian nature of the Lorentz anomaly, $W$ satisfies the Wess-Zumino consistency condition for Lorentz transformations on manifolds with a boundary once we add to it an appropriate boundary term,
\begin{equation}
	\delta_v W = \tilde{c} \int_{\mathcal{M}} \theta d\omega
	\qquad
	\delta_{\xi}W = 0\,.
\end{equation}
It is also straightforward to demonstrate that $[\delta_\xi,\delta_v]W = \delta_{[\xi,v]}W$ where $\delta_{[\xi,v]} = 0$. 

In the absence of boundaries Lorentz anomalies can be converted to Einstein (diffeomorphism) anomalies by adding appropritate local counterterms to the action. The technical term is that the gravitational anomaly is mixed between diffeomorphisms and local rotations. Indeed, following \cite{Bardeen:1984pm},  consider the following  expression:
\begin{equation}
\label{E:DeltaS}
	\Delta S = \tilde{c} \int_0^1 ds \int_{\mathcal{M}}   \hbox{Tr}\left( H d\omega(s)\right)
\end{equation}
where we have defined a vielbein $e(s)$ which interpolates between $e(0)^{a}{}_{\mu} = \delta^{a}{}_{\mu}$ and $e(1)^a{}_{\mu} = e^a{}_{\mu}$ where $e$ is the vielbein on $\mathcal{M}$. Treating $e(s)$ as a matrix-valued zero form, we have defined 
\begin{align}
\label{E:defHws}
	H = e(s)^{-1}  \frac{\partial}{\partial s}  e(s)\,,
	\quad
	\omega(s) = e(s)^{-1} \left(\omega \epsilon+ d \right) e(s)
\end{align}
where $\omega \epsilon = \omega_{\mu}dx^{\mu} \epsilon^{ab}$ is the spin connection one-form associated with $e^a{}_{\mu}$. Note that $\omega(1) = \Gamma$, where $\Gamma$ is the Christoffel connection associated with the vielbein $e$, and considered as a gauge transformation of $\omega$. 

Since $\Delta S$ is a functional of $e$ and $de$ then it is a trivial solution to the Wess-Zumino consistency conditions \eqref{E:WZconsistency}. Further, one can show that 
\begin{align}
\begin{split}
\label{E:deltaDeltaS}
	\delta_\theta \Delta S &= -\tilde{c}\int_{\mathcal{M}} \theta d\omega +\tilde{c} \int_0^1 ds \int_{\partial \mathcal{M} } \hbox{Tr} \left(H [\omega(s),\theta(s)] \right) \\
	\delta_{\xi} \Delta S &= \tilde{c}\int_{\mathcal{M}} \partial_{\mu}\xi^{\nu} d \Gamma^{\mu}{}_{\nu} +\tilde{c} \int_0^1 ds \int_{\partial \mathcal{M} } \hbox{Tr} \left(H [\omega(s),\xi(s)] \right) \,,
\end{split}
\end{align}
where
\begin{equation}
	\theta(s) \equiv e(s)^{-1} \theta \epsilon e(s)  + e(s)^{-1} \delta_\theta e(s)
	\quad
	\xi(s) = e(s)^{-1} \delta_{\xi} e(s)\,,
\end{equation}
and $[A,B]$ is the commutator. See \cite{bertlmann,Bardeen:1984pm}\footnote{To obtain \eqref{E:deltaDeltaS}, we found the following identities useful
\begin{align}
\begin{split}
	\delta_\theta \omega(s) &= d\theta(s)+[\omega(s) ,\theta(s)]\,,
	\qquad
	\delta_{\xi} \omega(s)  = d\xi(s) + [\omega(s),\,\xi(s)]\,,
	\qquad
	\delta_\theta H = [H,\,\theta(s)]+\frac{\partial \theta(s)}{\partial s} \\
	\delta_\xi H &= [H,\,x(s)]+\frac{\partial x(s)}{\partial s} \,,
	\qquad
	\frac{\partial \omega(s)}{\partial s} = dH-[H,\,\omega(s)]\,.
\end{split}
\end{align}	
} for an extensive discussion. 

As we have emphasized, $\Delta S$ is local in $e$ and $de$, so we may think of it as a contact term which may be added to $W$. Thus, if we define $\tilde{W} = W + \Delta S$ we obtain 
\begin{align}
\begin{split}
	\delta_{\xi} \tilde{W} &=\tilde{c} \int_{\mathcal{M}} \partial_{\mu}\xi^{\nu} d \Gamma^{\mu}{}_{\nu} +\tilde{c} \int_0^1 ds \int_{\partial \mathcal{M} } \hbox{Tr} \left(H\, [\omega(s),\xi(s)] \right) \,,
	\\
	\label{E:EinsteinAnomaly}
	\delta_\theta \tilde{W}  &= \tilde{c}\int_0^1 ds \int_{\partial \mathcal{M} } \hbox{Tr} \left(H\, [\omega(s),\theta(s)] \right)\,, \hspace{-.07in}
\end{split}
\end{align}
The bulk term on the right hand side of $\delta_{\xi}\tilde{W}$ is the standard expression for the Einstein anomaly \cite{Bardeen:1984pm,bertlmann} as expected. One may check that on its own, it does not satisfy the Wess-Zumino consistency condition \eqref{E:WZconsistency} on a manifold with a boundary. The boundary term associated with the Einstein variation of $\tilde{W}$ appearing on the far right of $\delta_{\xi}\tilde{W}$ precisely compensates for the bulk term's violation of the Wess-Zumino consistency condition, so that $\tilde{W}$ indeed satisfies \eqref{E:WZconsistency} More generally, $\delta_\theta\tilde{W}$ together with $\delta_{\xi}\tilde{W}$ ensure that \eqref{E:WZconsistency} is satisfied for both Lorentz and Einstein variations.

The main lesson we have learned from the analysis of the two-dimensional gravitational anomaly is that once a mixed anomaly satisfies the Wess-Zumino consistency condition in one frame, then shifting the anomaly to another frame by adding a local term to the quantum effective action will not lead to a violation of the Wess-Zumino consistency conditions.

We conclude this section with an analysis of conformal field theories (CFTs). (In the next Section we perform a complementary analysis in terms of two-point functions on the upper half-plane.) In a two-dimensional CFT the gravitational anomaly, whose strength is parameterized by $\tilde{c}$, is associated with an asymmetry in the left and right central charges,
\beq
	\tilde{c} = \frac{c_L - c_R}{96\pi}\,,
\eeq
whereas the total central charge is proportional to their sum,
\beq
	c = \frac{c_L + c_R}{24\pi}\,.
\eeq
When $c\neq 0$ two-dimensional CFT's have an anomaly under infinitesimal Weyl rescalings of the metric,
\beq
	\label{E:deltaSigma}
	\delta_{\sigma} g_{\mu\nu} = 2\sigma \, g_{\mu\nu}\,,
	\qquad
	\delta_{\sigma} e^a{}_{\mu} = \sigma\, e^a{}_{\mu}\,.
\eeq
The Weyl anomaly is Abelian and satisfies
\begin{equation}
	\delta_{\sigma_1}\sigma_2 = 0
	\qquad
	[\delta_{\sigma_1},\,\delta_{\sigma_2}]=0\,.
\end{equation}	
The Weyl anomaly on a manifold with boundary has long been known~\cite{polchinski1998string} to be
\beq
\label{E:Weyl}
\delta_{\sigma}W = - c \left( \int_{\M} \sigma\, d\omega + \int_{\partial \M}\sigma\, K\right)\,,
\eeq
with $K$ the extrinsic curvature one-form. In two dimensions the spin connection $\omega$ is related to the scalar curvature $R$ by
\beq
	d\omega = \frac{1}{2}d^2x \sqrt{g} R\,,
\eeq
which can be used to bring \eqref{E:Weyl} into a more canonical form. 

Let us now consider CFTs where both $c$ and $\tilde{c}$ are nonzero, meaning theories with both a Weyl and Lorentz anomaly. We denote the action of a joint infinitesimal Weyl scaling with parameter $\sigma$ and a local rotation with parameter $\theta$ by $\delta_v$, such that, e.g.,
\beq
\label{E:Weyl1}
	\delta_{v}\omega = d\theta + \star d\sigma\,.
\eeq
We note that
\begin{equation}
\label{E:Weyl2}
	\delta_{v}\theta = \delta_{v}\sigma=0
\end{equation}
and that $[\delta_{v_1},\,\delta_{v_2}]=0$. 

We start with manifolds without boundaries. In order to solve the Wess-Zumino consistency conditions for the Weyl anomaly we must include, at the very least, the bulk term in \eqref{E:Weyl}, $-c\int_{\mathcal{M}} \sigma d\omega$. Likewise, in order to solve the Wess-Zumino consistency condition for the gravitational anomaly we must include, at the very least, a bulk  term, $\int_{\mathcal{M}} \theta d\omega$ as in \eqref{E:LorentzAnomaly}. One can check that these two contributions alone do not satisfy the mixed Weyl-Lorentz Wess-Zumino consistency condition. To remedy this, we include an additional bulk term, $\int_{\M}d\sigma \wedge \star\omega$ such that
\beq
\label{E:CFTanomaly}
	\delta_{v}W =- c \int_{\M} \sigma\, d\omega + \tilde{c}\int_{\M} \left( \theta d\omega -d  \sigma \wedge \star\omega\right)\,.
\eeq
One can check that \eqref{E:CFTanomaly} is fully consistent with \eqref{E:WZconsistency}. The last term on the right hand side of \eqref{E:CFTanomaly} is associated with a mixed Lorentz-Weyl anomaly \cite{chamseddine1992}.

In the presence of a boundary the Lorentz anomaly is trivially consistent, and the Weyl anomaly can be made consistent by adding a boundary term proportional to the extrinsic curvature as in \eqref{E:Weyl}. With some prescience our candidate for $\delta_vW$ is
\begin{equation}
\label{E:candidate2d}
	\delta_{v}W \overset{?}{=} - c \left( \int_{\M} \sigma\, d\omega +\int_{\partial\mathcal{M}} \sigma K \right) 
	+ \tilde{c} \left( \int_{\M} \left( \theta d\omega - d\sigma \wedge \star \omega\right) + \int_{\partial \mathcal{M}} \theta K \right)  \,.
\end{equation}
Thus,
\beq
\label{E:2dCFTobstruction2}
[\delta_{v_1},\delta_{v_2}]W =- \tilde{c} \int_{\partial\M} (\sigma_2d\sigma_1 - \sigma_1d\sigma_2)\,.
\eeq
The only time-reversal-violating and diffeomorphism-invariant boundary terms that could be added to \eqref{E:candidate2d} in order to set $[\delta_{v_1},\,\delta_{v_2}]W = 0$ are
\beq
\label{E:nodeltaW}
	\int_{\partial \M} \theta \,K\,, \qquad \int_{\partial \M} \sigma \star\omega\,.
\eeq
It is straightforward to show that neither of these ensure that the Wess-Zumino consistency condition be satisfied.

At this point the careful reader may wonder whether it is possible to render the Lorentz-Weyl anomaly consistent by allowing for a diffeomorphism anomaly on the boundary.\footnote{A few comments are in order for the reader who is interested in this point. Suppose that it is the case that a diffeomorphism anomaly on the boundary renders the Lorentz-Weyl consistent. Because the Lorentz anomaly is consistent on its own, this would imply that two-dimensional systems can support a consistent diffeomorphism anomaly on their boundary. That is, this would imply the existence of a new $0+1$-dimensional diffeomorphism anomaly which can only live on the boundary of a $2d$ system. Requiring that such an anomaly follow from descent relations (as we have done implicitly in deriving \eqref{E:hatdQ0}) implies that it can not exist; there is no candidate three dimensional characteristic class which would lead to a one dimensional gravitational anomaly on a boundary.} Instead of taking this route we will show in Section \ref{S:2dCFT}, using conformal invariance and the Ward identities, that two dimensional CFT's on a half-plane do not allow for flavor or gravitational 't Hooft anomalies, Abelian or not.

\subsection{Four dimensions}
\label{SS:4d}

Our analysis of four-dimensional anomalies parallels that of the two-dimensional ones. The solution to the Wess-Zumino consistency condition for a cubed flavor anomaly  in the bulk is given by \eqref{E:vGsolution}
\begin{equation}
	\delta_{\Lambda} W = -c_A\int \hbox{Tr} \left(d\Lambda \left( F A - \frac{1}{2} A^3 \right) \right)\,.
\end{equation}
so that \eqref{E:Qsolution} takes the form
\begin{equation}
	[\delta_{\Lambda_1},\,\delta_{\Lambda_2}] W - \delta_{[\Lambda_1,\Lambda_2]} W =c_A \int_{\partial \mathcal{M}} \hbox{Tr} \left([d\Lambda_1,\,d\Lambda_2]A \right)
\end{equation}
which is non-trivial unless the symmetry is Abelian. Similarly, if the anomaly were mixed between an Abelian group and a non-Abelian one, then we may write
\begin{equation}
\label{E:mixedflavor}
	\delta_{v} W = -c_{an} \int \Lambda_a \hbox{Tr}(F_n\wedge F_n)
\end{equation}
where $\Lambda_a$ is an Abelian gauge transformation and $F_n$ is a non Abelian field strength. It is straightforward to check that \eqref{E:mixedflavor} satisfies \eqref{E:WZconsistency}.

In four dimensions there is no pure gravitational anomaly, but there is a mixed flavor-gravitational anomaly. The anomaly is mixed between a $U(1)$ flavor symmetry and either diffeomorphisms or (and) local Lorentz rotations. Placing the anomaly in the flavor sector one finds that 
\begin{align}
\begin{split}
\label{E:4d}
	\delta_{v} W =&  \tilde{c} \int_\M d^4x \sqrt{g} \,  \Lambda \epsilon^{\mu\nu\rho\sigma}R^{\alpha}{}_{\beta\mu\nu}R^{\beta}{}_{\alpha\rho\sigma} 
	 + \frac{a}{16\pi^2} \left(\int_{\mathcal{M}} d^4x \sqrt{g}\, \sigma E_4 - \int_{\partial \mathcal{M}} d^3x \sqrt{\gamma}\, \sigma Q_4 \right) \\
	&-\frac{c}{16\pi^2} \int_{\mathcal{M}}  d^4x \sqrt{g} \,\sigma W^2 +
	b_1  \int_{\partial \mathcal{M}} d^3x \sqrt{\gamma} \,\sigma K_1 
	+ b_2 \int_{\partial \mathcal{M}} d^3x \sqrt{\gamma} \,\sigma K_2 \,.
\end{split}
\end{align}
satisfies the Wess-Zumino consistency conditions.
Here $v$ specifies a $U(1)$ flavor transformation with parameter $\Lambda$, a diffeomorphism with parameter $\xi$ and Weyl rescaling with parameter $\sigma$. The variation \eqref{E:4d} satisfies the Wess-Zumino consistency condition in the presence of a boundary. The relevant transformation laws are identical to the ones appearing in equations \eqref{E:rules1}, \eqref{E:Lorentztransformation}, \eqref{E:geometry1}, \eqref{E:geometry2} and generalize \eqref{E:Weyl1} and \eqref{E:Weyl2} of the previous section. We have collected the relevant equations in appendix \ref{A:alltransformations} for convenience. The coefficient $\tilde{c}$ in \eqref{E:4d} characterizes the strength of the mixed gauge-gravitational anomaly and the coefficients $c$ and $a$ characterize the bulk  Weyl  anomaly (so that $W^2$ represents the Weyl tensor squared and $E_4$ the four dimensional Euler density). In addition to $c$ and $a$ there exist boundary central charges $b_1$ and $b_2$ which satisfy the Wess-Zumino consistency condition on the boundary independent of $a$ and $c$. We refer the reader to \cite{Herzog:2015ioa} for a detailed exposition and precise definitions of $W^2$, $E_4$, $Q_4$, $K_1$ and $K_2$ (see also~\cite{Jensen:2015swa,Fursaev:2015wpa,Solodukhin:2015eca}).

Since the bulk terms in \eqref{E:4d} are the most general ones compatible with Weyl and mixed anomalies, and since $\delta_{v}W$ satisfies the Wess-Zumino consistency condition, then it will also satisfy it if the anomaly is shifted to the Einstein or Lorentz sector. Thus, as far as Wess-Zumino consistency is concerned, four-dimensional theories with mixed anomalies may be put on manifolds with boundaries.

\subsection{Higher dimensions}
\label{SS:higherd}

As we have demonstrated, anomalous theories can be consistently placed on a manifold with boundary only if the anomaly is pure Abelian or is mixed with an Abelian symmetry. Since the gravitational anomaly in $d>2$ spacetime dimensions is non-Abelian, only anomaly polynomials of the form 
\begin{equation}
\label{E:Finalresult}
	P = F \wedge \ldots
\end{equation}
with $F=dA$ an Abelian field strength are allowed 
on spaces with a boundary. For instance, six-dimensional theories with pure gravitational anomalies characterized by an anomaly polynomial $P = \hbox{Tr} \left(Riemm^4\right)$ (with $Riemm$ representing the Riemann curvature two-form) are not consistent, but mixed  flavor-gravitational anomalies with $P=F\wedge F \wedge \hbox{Tr} \left(Riemm^2\right)$ (with $F$ an Abelian field strength) are. 

It is interesting to note that recent reductions of six dimensional theories on punctured Riemann surfaces exhibit precisely such a feature---the anomalous non-Abelian symmetries of these theories are broken so that the anomaly polynomial is of the form \eqref{E:Finalresult} \cite{Razamat:2016dpl,ShlomoNew}.

\section{Two dimensional CFT}
\label{S:2dCFT}

In Section~\ref{SS:2d} we have argued that Wess-Zumino consistency condition forbids non-Abelian anomalies on a manifold with boundary, and provided evidence that the gravitational anomaly of two-dimensional CFT is also inconsistent. In this Section we perform a complementary analysis in terms of the two-point functions of the stress tensor and flavor currents of a two-dimensional CFT on half-space, i.e. of a boundary CFT. We find that the boundary does not allow for a gravitational anomaly, nor flavor anomalies of any kind. We assume locality, unitarity, conformal invariance, and the Ward identities for the stress tensor and flavor currents. The authors of~\cite{watanabe,simeon} have also shown that two-dimensional boundary CFTs do not have gravitational or flavor anomalies within the boundary state formalism.

Consider a local two dimensional CFT on a Euclidean background specified by Cartesian coordinates $x$ and $y$ such that $y \geq 0$ and $x \in \mathbb{R}$. The boundary breaks the global conformal group from $SL(2,\mathbb{C})/\mathbb{Z}_2$ to $SL(2,\mathbb{R})/\mathbb{Z}_2$. Indeed, defining $z=x+iy$, in the absence of a boundary the conformal group acts on $z$ and $\bar{z}$ via 
\begin{equation}
\label{E:global2dcft}
	z \to \frac{a z + b}{c z +d}\,,
	\qquad
	\bar{z}  \to \frac{\bar{a} \bar{z} + \bar{b}}{\bar{c} \bar{z} + \bar{d}}\,. 
\end{equation}
In the presence of a boundary (which in the new coordinate system is located along $z=\bar{z}$) the subgroup of \eqref{E:global2dcft} which preserves the boundary is $SL(2,\mathbb{R})/\mathbb{Z}_2$.

The Ward identities for the stress tensor in the presence of a boundary are
\begin{equation}
\label{E:Ward2d}
	T^{\mu}{}_{\mu} = 0\,, \qquad \partial_{\nu}T^{\mu\nu} = n^{\mu} \mathcal{D} \delta(x_{\bot})\,.
\end{equation}
Here $n^{\mu}$ is a normal to the surface and $\mathcal{D}$ is referred to as the displacement operator. It captures information regarding non-conservation of momentum through the boundary due to loss of translation invariance. Formally, one can obtain \eqref{E:Ward2d} by considering the variation of the generating function of connected correlators under an infinitesimal coordinate transformation. The generating function is a functional of the metric tensor and other sources and in the presence of a boundary it is also a functional of the embedding function of the boundary. The displacement operator is the operator conjugate to this embedding function. We refer the reader to \cite{McAvity:1993ue,Liendo:2012hy,Jensen:2015swa,Herzog:2017xha,Herzog:2017kkj} for various discussions.

We note that equation \eqref{E:Ward2d} is valid in the presence of conformal or gravitational anomalies. Since the background metric and boundary are flat the Riemannian and extrinsic curvature vanish and there are no additional contributions to \eqref{E:Ward2d}.
Further, since both the $x,\,y$ coordinates and the $z,\,\bar{z}$ coordinates have vanishing Christoffel connection the non tensorial properties of $T^{\mu\nu}$ don't modify \eqref{E:Ward2d} either.  We also note that \eqref{E:Ward2d} will not be modified by adding conformal boundary degrees of freedom to $y=0$; in $0+1$ dimensions, $T^{\mu\nu}$ has one component which must be set to zero due to conformal invariance.\footnote{One may wonder whether there exist Weyl, Lorentz, and (or) diffeomorphism breaking boundary terms which modify the Ward identities even in the absence of curvature terms. As far as we know, such terms have never been observed in the literature. Using a canonical scaling dimension for the stress tensor one may consider all possible boundary modifications to \eqref{E:Ward2d}, an example of which would be $T^{\mu}{}_{\mu} = b_1 \delta'(y)$. One may check that the conclusion of the ensuing analysis will remain unchanged even in the presence of such terms.}

One may still attempt to argue that $T^{\mu\nu}$ may have boundary contributions. To see that this is not the case let us consider a stress tensor of the form:\footnote{One may allow for more general distributional terms, e.g., $y^{-n}\delta(y)$. Our argument still goes through, as only the distributions shown are related by the Ward identities to each other and $T^{(0)xy}$ at the boundary.}
\begin{equation}
\label{E:boundaryterms}
	T^{\mu\nu} = T^{(0)\mu\nu} + \sum_{n=1}^2 T^{(n)\mu\nu} \partial_y^{n-1}\delta(y)  \,.
\end{equation}
The reason our series truncates at $n=2$ is that the unitarity bound for a $0+1$ dimensional conformal quantum mechanics is $-1/2$. Using
\begin{equation}
	\lim_{\epsilon \to 0} \int_{-\epsilon}^{\epsilon} y^m \partial_{\mu} T^{\mu\nu} dy = \delta^{\nu}_y \mathcal{D} \,\delta^{m0}\,,
\end{equation}
for $m=0\ldots,2$ we find
\begin{equation}
	T^{(m)y\nu} = 0\,,
\end{equation}
which together with the trace Ward identity implies that $T^{(m)\mu\nu}=0$ for $m>0$. Thus, there can be no boundary contributions to the stress tensor. In addition, we have
\begin{equation}
\label{E:noflux}
	T^{(0)xy}\big|_{y=0} = 0
\end{equation}
which would imply that there is no (Euclidean) flux of energy through the boundary.

We are now prepared to study the implications of \eqref{E:Ward2d}. Going to the $z,\,\bar{z}$ coordinate system we find that away from the boundary \eqref{E:Ward2d} implies the standard holomorphic decomposition of the stress tensor,
\begin{equation}
	T_{z\bar{z}}=0
	\qquad
	T_{zz}=T(z)
	\qquad
	T_{\bar{z}\bar{z}}=\bar{T}(\bar{z})\,.
\end{equation}
Thus, given the $SL(2,\mathbb{R})/\mathbb{Z}_2$ symmetry of the theory we find
\begin{equation}
\label{E:correlators}
	\langle T(z) T(z') \rangle = \frac{c_L}{8\pi^2(z-z')^4}
	\qquad
	\langle \bar{T}(\bar{z}) \bar{T}(\bar{z}') \rangle =  \frac{c_R}{8\pi^2(\bar{z}-\bar{z}')^4}\,.
\end{equation}
The no flux condition \eqref{E:noflux} implies that the operator identity 
\begin{equation}
\label{E:noflux2}
	T\big|_{y=0}=\bar{T}\big|_{y=0}
\end{equation}
should hold at the boundary. Imposing \eqref{E:noflux2} on \eqref{E:correlators} we find that we must set $c_L=c_R$.

A similar argument constrains flavor anomalies. Consider a two-dimensional boundary CFT with flavor current $J_{\mu}^a$ where $a$ represents a flavor index associated with a flavor symmetry $G$. Away from other insertions, the currents satisfy
\beq
\label{E:currentconservation}
	\partial_{\mu}J^{\mu} = 0\,.
\eeq
In the absence of a boundary the full $SL(2,\mathbb{C})$ conformal symmetry implies that the components $J_z$ and $J_{\bar{z}}$ are separately conserved. The flavor symmetry is enhanced to $G\times G$, and is characterized by Kac-Moody levels $k_L$ and $k_R$. As we will show shortly, when there is a boundary the current still decomposes into holomorphic and anti-holomorphic sectors, but when $k_L=k_R$ the boundary breaks the symmetry down to the diagonal subgroup $G$.

As in our analysis of the stress tensor, suppose that the current has distributional terms of the form
\beq
J^{\mu} = J^{(0)\mu} + J^{(1)\mu}\delta(y)\,.
\eeq
Integrating the Ward identity over the interval $y\in [-\epsilon,\epsilon]$ and taking $\epsilon\to 0$ we find that
\beq
J^{(0)y}\big|_{y=0} = \partial_x J^{(1)x}\,, \qquad J^{(1)y} = 0\,.
\eeq
The underlying conformal invariance implies that under an $SL(2,\mathbb{R})/\mathbb{Z}_2$ transformation, the current transforms as
\beq
\label{E:transformationOfJ}
	J_z(z,\bar{z})\to (c z+d)^2 J_z(z,\bar{z})\,, \qquad J_{\bar{z}}(z,\bar{z}) \to (c \bar{z}+d)^2 J_{\bar{z}}(z,\bar{z})\,.
\eeq
If $J^{(1)x}$ is nonzero, then it is a dimension-0 boundary operator, with
\beq
	\langle J^{(1)x}(x)J^{(1)x}(x')\rangle \propto \ln (x-x')^2\,.
\eeq
It is straightforward to check that $J^{(1)\mu}\delta(y)$ does not transform as in~\eqref{E:transformationOfJ} implying that
\beq
J^{(1)x} = 0\,,
\eeq
so that the current has no distributional term. The Ward identity then sets
\beq
\label{E:noJflux}
J^{(0)y}\big|_{y=0} = 0\,.
\eeq

Next consider the two-point function of the current, $\langle J_{\mu}(z,\bar{z})J_{\nu}(z',\bar{z}')\rangle$. There is a single conformally invariant cross-ratio formed by the two insertions~\cite{McAvity:1992fq}
\beq
v = \frac{|z-z'|^2}{|z-\bar{z}'|^2}\,.
\eeq
The boundary is located at $v= 1$ . Conformal invariance constrains the two-point function up to three free functions of $v$,
\begin{align}
\begin{split}
\langle J^a_{z}(z,\bar{z})J^b_{z}(z',\bar{z}')\rangle &= \frac{g_1(v) \delta^{ab}}{(z-z')^2}\,, 
\\
\langle J^a_{\bar{z}}(z,\bar{z})J^b_{\bar{z}}(z',\bar{z}')\rangle &= \frac{g_2(v) \delta^{ab}}{(\bar{z}-\bar{z}')^2}\,, 
\\
\langle J^a_{z}(z,\bar{z})J^b_{\bar{z}}(z',\bar{z}')\rangle & = \frac{g_3(v) \delta^{ab}}{(z-\bar{z}')^2}\,.
\end{split}
\end{align}
Current conservation implies that the $g_i(v)$ are constants, and consequently the holomorphic and anti-holomorphic components of the current are separately conserved,
\beq
J_z = J(z)\,, \qquad J_{\bar{z}} = \bar{J}(\bar{z})\,.
\eeq
The two-point function of $J$ and $\bar{J}$ then take the same form as in a CFT on the plane,
\beq
\label{E:2pointJ}
\langle J^a(z)J^b(z')\rangle = \frac{k_L \delta^{ab}}{\pi^2 (z-z')^2}\,, \qquad \langle \bar{J}^{a}(\bar{z})\bar{J}^{b}(\bar{z}')\rangle = \frac{k_R \delta^{ab}}{\pi^2(\bar{z}-\bar{z}')^2}\,.
\eeq
The no-flux condition~\eqref{E:noJflux} implies the operator identity
\beq
J\big|_{y=0} = \bar{J}\big|_{y=0}\,,
\eeq
which implies that the Kac-Moody levels must satisfy $k_L = k_R$.

In the absence of a boundary, the holomorphic and anti-holomorphic currents are separately conserved, and the flavor symmetry $G$ is enhanced to a $G\times G$ flavor symmetry. Said another way, both $J^{\mu}$ and $\epsilon^{\mu\nu}J_{\nu}$ are conserved. In the usual language $J^{\mu}$ is the vector current and $\epsilon^{\mu\nu}J_{\nu}$ is the axial current. In the presence of a boundary both the vector current and axial current satisfy \eqref{E:currentconservation} in the bulk. But only the vector current satisfies it on the boundary, viz., \eqref{E:noJflux}. Thus, the axial current is not conserved. Consequently the boundary breaks the $G\times G$ flavor symmetry down to the diagonal vector-like subgroup $G$. The boundary explicitly breaks the axial symmetry but leaves the vector symmetry intact, and thus the boundary CFT has no flavor anomaly.

Putting the pieces together, we learn that a $G\times G$ flavor symmetry is broken by the boundary to a vector-like, non-anomalous subgroup. This can only be done if
\beq
k_L = k_R\,.
\eeq
Put this way there is a clear analogy with the stress tensor. The boundary also breaks the underlying $SL(2;\mathbb{C})/\mathbb{Z}_2\sim SL(2,\mathbb{R})/\mathbb{Z}_2\times SL(2,\mathbb{R})/\mathbb{Z}_2$ conformal symmetry down to the diagonal $SL(2,\mathbb{R})/\mathbb{Z}_2$ subgroup, and this can only be done if $c_L = c_R$.

\section{Discussion}
\label{S:discussion}

In this work we have argued that the Wess-Zumino consistency condition prohibits the existence of 't Hooft anomalies on manifolds with boundary at least when the anomalies are not of the form given in \eqref{E:Finalresult}. This observation is in line with earlier works \cite{Luckock:1990xr} which suggested that free chiral fermions can not be canonically quantized on manifolds with boundaries---at least in the absence of bulk to boundary interactions. Indeed, we also directly showed that apart from the Weyl anomaly, two-dimensional boundary CFTs do not possess 't Hooft anomalies. In what follows we discuss these results and their ramifications in some detail.

In more than two dimensions we have exhibited an obstruction to placing anomalous theories on a manifold with boundary but we have not discussed the mechanism for the obstruction. When constructing the partition function for a theory described by a Lagrangian, one must integrate over quantum fields subject to boundary conditions. We interpret our obstruction as the statement that, when a theory has a non-Abelian anomaly, these boundary conditions necessarily break the non-Abelian symmetry down to a subgroup with at most an Abelian factor, and, in two-dimensions, the boundary conditions completely break the anomalous global symmetries. Similar arguments were made in \cite{Razamat:2016dpl,ShlomoNew} to explain the breaking of non-Abelian flavor symmetries to an Abelian subgroup for six dimensional superconformal field theories (SCFTs) on punctured Riemann surfaces.

Our results raise interesting questions regarding renormalization group (RG) flow. On the one hand, we have demonstrated that two-dimensional boundary CFTs do not possess 't Hooft anomalies. On the other, we are unable to rule out the possibility that higher-dimensional theories have Abelian anomalies, and indeed, there appear to be examples of such theories in $d=4$ descending from the E-string theory~\cite{Razamat:2016dpl,ShlomoNew}. Consider putting such a theory on a space of the form $\mathbb{H}^2\times \M$, where $\mathbb{H}^2$ is the upper half-plane and $\M$ is a Riemann surface threaded with flux for the anomalous Abelian global symmetry. The low-energy theory on $\mathbb{H}^2$ necessarily has an 't Hooft anomaly. But then it is not clear what sort of theory lives at the endpoint of the flow; perhaps the RG flow reaches a limit cycle in the infrared. Relatedly, scale invariance implies conformal invariance in two dimensions, but to our knowledge it is not known if this remains true on $\mathbb{H}^2$. So another possibility is that the endpoint is scale-but-not-conformally invariant. Yet another is that locality is somehow broken in the infrared, or perhaps even the starting point is ill-founded. Given that our proof relies on the Wess-Zumino consistency condition which is known to be ill-suited for handling Abelian anomalies, and that canonical quantization of free fields (and our results for two dimensionsal CFTs) are incompatible with possessing any sort of chiral matter altogether, we find it most likely that theories with flavor or gravitational anomalies on manifolds with boundaries are inconsistent, regardless of the group structure of the anomalies.

In this work we limited our analysis to continuous global symmetries. It would be very interesting to extend our analysis to anomalous discrete symmetries.

Finally, our result for two-dimensional CFTs provides some insight into the study of entanglement entropy of CFTs with a gravitational anomaly \cite{Castro:2014tta,Iqbal:2015vka,Nishioka:2015uka,Azeyanagi:2015uoa}. When computing entanglement entropy one implicitly constructs a boundary, the entangling surface, separating two spatial regions $A$ and $\bar{A}$. To construct the theory on $A$, one implicitly imposes boundary conditions on its boundary~\cite{Ohmori:2014eia}. In the current context, constructing such a boundary implies the breaking of conformal symmetry. Thus, it may be the case that naive use of the replica trick in computing entanglement entropy is inappropriate when $c_L\neq c_R$.\footnote{Similar conclusions were reached in~\cite{watanabe,simeon}, and we are indebted to S.~Hellerman for many lively and insightful discussions on this point, especially for emphasizing the importance of~\cite{Ohmori:2014eia}.} Of course, since the results of \cite{Iqbal:2015vka,Nishioka:2015uka} are valid for conformal as well as non-conformal theories, the significance of this observation on the general validity of the aforementioned computations is not clear. We hope to return to this issue in the future.

\section*{Acknowledgements.} We would like to thank J.~David, S.~Hellerman, C.~Herzog, S.~Razamat, and A.~Schwimmer for many enlightening conversations. KJ was supported in part by the US Department of Energy under Grant No. DE-SC0013682. ES and AY were supported by the ISF under grant number 1989/14.

\begin{appendix}
\section{Transformation rules}
\label{A:alltransformations}

For the readers convenience we provide a concise collection of the transformation laws for all external sources appearing in this work
\begin{align}
\begin{split}
	\delta_{v} g_{\mu\nu}  =&  \partial_{\mu}\xi^{\rho} g_{\rho\nu} + \partial_{\nu}\xi^{\rho}g_{\mu\rho} + 2\sigma \,g_{\mu\nu} \,,
\\
	\delta_{v}e^a_{\mu} =& \sigma \, e^A_{\mu} - \theta^a{}_b  e^b_{\mu} + \partial_{\mu}\xi^{\nu} e^a_{\nu}\,,
\\
	\delta_{v} A_{\mu} =& \partial_{\mu}\xi^{\nu}A_{\nu} + \partial_{\mu}\Lambda + [A_{\mu},\Lambda]\,,
\\
	\delta_{v} \omega_{\mu}{}^a{}_b = & \omega_{\mu}{}^a{}_c \theta^c{}_b - \theta^a{}_c \omega_{\mu}{}^c{}_b + \partial_{\mu} \theta^{a}{}_b  + (\partial_{\tau}\sigma) g^{\tau\rho} g_{\mu\nu} e^{a}{}_{\rho} e^{\nu}{}_{b} - (\partial_{\nu} \sigma) e^a{}_{\mu} e^{\nu}{}_{b}
\\
	\delta_{v}\Gamma^{\mu}{}_{\nu\rho} =& - \partial_{\alpha}\xi^{\mu} \Gamma^{\alpha}{}_{\nu\rho} + \partial_{\nu}\xi^{\alpha}\Gamma^{\mu}{}_{\alpha\rho} + \partial_{\rho} \xi^{\alpha}\Gamma^{\mu}{}_{\nu\alpha}  +\partial_{\nu}\partial_{\rho}\xi^{\mu} + \delta^{\mu}_{\nu}\partial_{\rho}\sigma + \delta^{\mu}_{\rho}\partial_{\nu}\sigma - g_{\nu\rho}\partial^{\mu} \sigma  \,,
\end{split}
\end{align}
Note that
\begin{equation}
	\delta_v \partial_{\mu} = \partial_{\mu}\delta_v + \left( \partial_{\mu}\xi^{\alpha} \right) \partial_{\alpha}  \,.
\end{equation}

Next we provide transformation laws for all the gauge parameters.
\begin{align}
\begin{split}
\delta_{v_1} \xi_2^{\mu}& = - \xi^{\nu}_2 \partial_{\nu} \xi_1^{\mu}\,,
\qquad
\delta_{v_1} \sigma_2 = 0\,,
\\
\delta_{v_1} \theta_2 & = 0 \,,
\hspace{.78in}
\delta_{v_1} \Lambda_2   = 0\,.
\end{split}
\end{align}

Finally, we provide the full expression for the transformation of commutators,
\begin{align}
\begin{split}
	\delta_{[\xi_1,\xi_2]} A_{\mu} &=\Big(\partial_{\mu}\left(\xi_1^{\rho}\partial_{\rho}\xi_2^{\nu} - \xi_2^{\rho}\partial_{\rho}\xi_1^{\nu} \right)\Big)A_{\nu} \,,\\ 
	\delta_{[\Lambda_1,\Lambda_2]} A & = d [\Lambda_1,\,\Lambda_2] + [A,[\Lambda_1,\,\Lambda_2]] \,, \\
	\delta_{[\theta_1,\theta_2]} V_a &= V_b ([\theta_1,\theta_2])^b{}_a\,.
\end{split}
\end{align}
All other transformations of this type vanish.

\section{Canonical (non-)quantization of chiral matter}
\label{A:canonical}

As a prelude let us consider a massless scalar field $\phi$ on $\mathbb{R}^{d-1,1}$,
\begin{equation}
\label{E:scalaraction}
	S = \frac{1}{2}\int d^dx(\partial\phi)^2\,.
\end{equation}
Decomposing $\phi$ into Fourier modes 
\begin{equation}
\label{E:modes}
	\phi(t,\vec{x}) = \frac{1}{(2\pi)^{d/2}} \int d^{d-1}k\, \hat{\phi}(t,\vec{k}) e^{i \vec{k} \cdot \vec{x}} \,,
\end{equation}
we find that the equations of motion on $\mathbb{R}^{d-1,1}$ imply
\begin{equation}
\label{E:modesol}
	\hat{\phi}(t,\vec{k}) \propto e^{\pm i t |\vec{k}|}
\end{equation}
One can then proceed with canonical quantization, imposing canonical commutation relations between $\phi$ and its conjugate momentum.

Instead of considering the scalar field on $\mathbb{R}^{d-1,1}$ let us consider the theory generated by \eqref{E:scalaraction} place on $\mathbb{R}^{d-2,1}\times \mathbb{H}$ where $\mathbb{H}$ denotes a semi-infinite interval. Equations \eqref{E:modes} and \eqref{E:modesol} go through unchanged. However, in order to ensure a well-defined variational principle we need to make sure that
\begin{equation}
\label{E:scalarBC}
	\int_{\mathbb{R}^{d-2,1}}  n^{\mu}  \delta {\phi} \, \partial_{\mu} \phi = 0\,.
\end{equation}
where $\mathbb{R}^{d-2,1}$ is the boundary of $\mathbb{R}^{d-2,1}\times \mathbb{H}$ and $n^{\mu}$ is a normal to the boundary. In what follows we will use a coordinate system where $n_{\mu}dx^{\mu} = dz$ and such that the boundary is located at $z=0$. Thus, \eqref{E:scalarBC} implies that $\phi$ must satisfy either Dirichlet or Neumann boundary conditions along $z=0$.
Imposing, e.g., Dirichlet boundary conditions on the mode expansion \eqref{E:modes} and \eqref{E:modesol} we find
\begin{equation}
	\phi(t,\vec{x}) = \frac{1}{(2\pi)^{d/2}} \int d^{d-1}k\,\left( \hat{\phi}_+(k) e^{i (|\vec{k}| t + \vec{k}\cdot \vec{x}_{\bot} )}\sin(k_z z) + \hat{\phi}_-(k) e^{i (-|\vec{k}| t + \vec{k}\cdot \vec{x}_{\bot} )}\sin(k_z z)\right)
\end{equation}
where now $x_{\bot}$ includes all spatial coordinates transverse to $z$ and $k_z = k^{\mu}n_{\mu}$. We may once again promote the $\phi_{\pm}$ to operators and proceed with canonical quantization.

We note that we have imposed the simplest type of boundary conditions possible. In practice it is possible to add boundary terms which enforce arbitrary values of $\phi$ on the boundary as in \cite{McAvity:1992fq}. Or, to add boundary degrees of freedom to the action \eqref{E:scalaraction} so that the boundary values of $\phi$ will be determined dynamically as in \cite{Herzog:2017xha}. While interesting, we will not discuss these boundary conditions further. Instead we turn our attention to fermions.

Let us consider $d=2n$ dimensional massless Dirac fermions
\begin{equation}
\label{E:Dirac}
	S = i\int d^dx \bar{\psi} \slashed{D} \psi \,.
\end{equation}
Requiring a well-defined variational principle implies that 
\begin{equation}
\label{E:variational}
	i\int_{\partial \mathcal{M}}d^{d-1}x\, \left( (\delta \bar{\psi}) \slashed{n} \psi +\bar{\psi}\slashed{n}\delta\psi\right) = 0\,,
\end{equation}
where $\slashed{n} = n^{\mu}\gamma_{\mu}$ and $n^{\mu}$ is the normal to the boundary. Note that  in order for the Dirac operator to be self adjoint we must have
\begin{equation}
\label{E:selfadjoint}
	\int_{\partial \mathcal{M}}d^{d-1}x\, \bar{\psi}_2 \slashed{n} \psi_1  = 0\,,
\end{equation}
which implies \eqref{E:variational}. Since the Dirac equation is first order in derivatives, imposing Dirichlet boundary conditions on $\partial \mathcal{M}$ completely fixes the solution in terms of the boundary conditions. Thus, one can not canonically quantize the free Dirac fermion of \eqref{E:Dirac} upon imposing Dirichlet boundary conditions on $\psi$. 

The problem of quantizing a free fermion on a manifold with a boundary is well known and has been discussed in detail in \cite{Luckock:1990xr}. Instead of imposing Dirichlet boundary conditions on all the eigenmodes one imposes boundary conditions on half of them. For completeness we will rederive this result. Instead of imposing Dirichlet boundary conditions on $\psi$ one may impose Dirichlet boundary conditions on a subset of its components. Indeed let us consider the projections
\begin{equation}
	P_{\pm} = \frac{1}{2} \left(1 \pm \chi \right)\,.
\end{equation}
We now impose
\begin{subequations}
\label{E:DiracBC} 
\begin{equation}
	P_{+} \psi \big|_{\partial \mathcal{M}} = 0
\end{equation}
which implies
\begin{equation}
	\bar{\psi} \bar{P}_{+}  \big|_{\partial \mathcal{M}} = 0
\end{equation}
\end{subequations}
where we have defined 
$
	\bar{P}_{\pm} = \frac{1}{2} \left(1 \pm \bar{\chi}\right)
$
with $\bar{\chi} = \gamma^0 \chi^{\dagger} \gamma^0$. We now require that the projection be such that \eqref{E:selfadjoint} take the form
\begin{equation}
\label{E:selfadjoint2}
	\bar{P}_- \slashed{n} P_- = 0\,.
\end{equation}

Demanding that $P_{\pm}$ and $\bar{P}_{\pm}$ are projections and that \eqref{E:selfadjoint} be satisfied implies that
\begin{equation}
\label{E:someconditions}
	\chi^2 = 1
	\qquad
	\bar{\chi}^2 = 1
	\qquad
	\slashed{n} \chi = - \bar{\chi} \slashed{n}
\end{equation}
We now impose the additional requirement that $P_{\pm}$ and $\bar{P}_{\pm}$ be invariant under Lorentz transformations in the directions orthogonal to the boundary. This last condition implies that $\chi$ can depend only on $\slashed{n}$ and the only available Lorentz scalar $\gamma_5$. Imposing \eqref{E:someconditions} gives us
\begin{equation}
\label{E:chival}
	\chi = i \slashed{n} f(i \gamma_5)\,,
\end{equation}
where $f$ is a real function. Consequently
\beq
\label{E:chiACgamma5}
\{\chi,\gamma_5\} = 0\,.
\eeq
The interested reader may refer to \cite{Marachevsky:2003zb} for the explicit form of $\chi$ for boundary conditions which are compatible with the MIT bag model.

We may now address the main problem we are interested in, imposing boundary conditions on free chiral fermions. Left and right-handed Weyl fermions are eigenspinors of $\gamma_5$, and so Eq.~\eqref{E:chiACgamma5} implies that $\chi$ flips a left-handed fermion into a right-handed one and vice versa. Thus, one can not impose the Dirichlet boundary conditions \eqref{E:DiracBC} on chiral fermions. A similar analysis follows for $N_R$ right handed fermions and $N_L$ left handed fermions. In order to implement the boundary conditions \eqref{E:DiracBC} one must set $N_R=N_L$.

\end{appendix}


\bibliographystyle{JHEP}
\bibliography{refs}

\end{document}